%% file: PolarAutomorphisms.tex
\documentclass[conference]{IEEEtran}

\renewcommand\IEEEkeywordsname{Index Terms}
\usepackage{graphicx}
\usepackage{subcaption}
\usepackage{tikz,xparse}
\usetikzlibrary{dsp,chains}
\usetikzlibrary{matrix}
\usetikzlibrary{spy}
\usepackage{mathptmx}
\usepackage{verbatim}
\usepackage{calc}
\usepackage{ifthen}
\usepackage{xifthen}
\usepackage{cancel}
\usepackage{bm}
\usepackage{verbatim}
\usepackage{multirow}
\usepackage{cite}

\usepackage[nolist]{acronym} 
\usepackage{pgfplots}
\usetikzlibrary{arrows,shapes,graphs,graphs.standard,quotes,decorations.markings}
\usetikzlibrary{matrix}
\pgfplotsset{compat=newest}

\usepackage[hyphens]{url}

\usepackage[bookmarks=false]{hyperref}
\usepackage{units}
\usepackage{amsmath, amsbsy, amssymb, latexsym }
\hypersetup{bookmarksdepth=-2}
\usepackage{comment}
\usepackage[utf8]{inputenc}
\usepackage{xcolor}
\usepackage{enumitem}
\usepackage[linesnumbered,vlined]{algorithm2e}
\usepackage{algpseudocode}

\usepackage{etoolbox} 

\SetAlCapNameFnt{\footnotesize}
\SetAlCapFnt{\footnotesize}

\tikzset{>=latex}

\DeclareMathOperator*\Stab{Stab}
\DeclareMathOperator*\ind{ind}
\DeclareMathOperator*\ord{ord}

\newcommand{\qed}{\hfill\blacksquare}


\input{corporateColours.tex}
\colorlet{Mycolor1}{green!10!orange!90!}
\input{nodesAndStyles.tex}

\IEEEoverridecommandlockouts

\input{macros}		

\def \negvspaces{1} 
\def \extended{1} 

\begin{document}
\begin{NoHyper}
\title{On the Automorphism Group of Polar Codes}

\author{\IEEEauthorblockN{Marvin Geiselhart, Ahmed Elkelesh, Moustafa Ebada, Sebastian Cammerer and Stephan ten Brink} \thanks{The authors would like to thank Florian Euchner for his help with proving Theorem 1 by proposing the \emph{crazy paternoster algorithm}.\newline\ifdef{\extended}{}{An extended version of this paper is provided online (arXiv:2101.09679) \cite{extended_arxiv}.}}
	\IEEEauthorblockA{
		Institute of Telecommunications, Pfaffenwaldring 47, University of  Stuttgart, 70569 Stuttgart, Germany 
		\\\{geiselhart,elkelesh,ebada,cammerer,tenbrink\}@inue.uni-stuttgart.de
	}
}

\makeatletter
\patchcmd{\@maketitle}
{\addvspace{0.5\baselineskip}\egroup}
{\addvspace{-0.6\baselineskip}\egroup}
{}
{}
\makeatother

\maketitle
\begin{acronym}
 \acro{ECC}{error-correcting code}
 \acro{HDD}{hard decision decoding}
 \acro{SDD}{soft decision decoding}
 \acro{ML}{maximum likelihood}
 \acro{GPU}{graphical processing unit}
 \acro{BP}{belief propagation}
 \acro{BPL}{belief propagation list}
 \acro{CA-BPL}{CRC-aided belief propagation list}
 \acro{LDPC}{low-density parity-check}
 \acro{HDPC}{high density parity-check}
 \acro{BER}{bit error rate}
 \acro{SNR}{signal-to-noise-ratio}
 \acro{BPSK}{binary phase shift keying}
 \acro{BCJR}{Bahl-Cocke-Jelinek-Raviv}
 \acro{AWGN}{additive white Gaussian noise}
 \acro{MSE}{mean squared error}
 \acro{LLR}{log-likelihood ratio}
 \acro{MAP}{maximum a posteriori}
 \acro{NE}{normalized error}
 \acro{BLER}{block error rate}
 \acro{PE}{processing element}
 \acro{SCL}{successive cancellation list}
 \acro{SC}{successive cancellation}
 \acro{BI-DMC}{Binary Input Discrete Memoryless Channel}
 \acro{CRC}{cyclic redundancy check}
 \acro{CA-SCL}{CRC-aided successive cancellation list}
 \acro{BEC}{Binary Erasure Channel}
 \acro{BSC}{Binary Symmetric Channel}
 \acro{BCH}{Bose-Chaudhuri-Hocquenghem}
 \acro{RM}{Reed--Muller}
 \acro{RS}{Reed-Solomon}
 \acro{SISO}{soft-in/soft-out}
 \acro{PSCL}{partitioned successive cancellation list}
 \acro{SPA}{sum product algorithm}
 \acro{LFSR}{linear feedback shift register}
 \acro{3GPP}{3rd Generation Partnership Project }
 \acro{eMBB}{enhanced Mobile Broadband}
 \acro{CN}{check node}
 \acro{VN}{variable node}
 \acro{PC}{parity-check}
 \acro{GenAlg}{Genetic Algorithm}
 \acro{AI}{Artificial Intelligence}
 \acro{MC}{Monte Carlo}
 \acro{CSI}{Channel State Information}
 \acro{FG}{factor graph}
 \acro{URLLC}{ultra-reliable low-latency communications}
 \acro{OSD}{ordered statistic decoding}
 \acro{LTA}{lower-triangular affine group}
 \acro{GA}{general affine group}
 \acro{BLTA}{block lower-triangular affine group}
 \acro{URLLC}{ultra-reliable low-latency communications}
 \acro{DMC}{discrete memoryless channel}
 \acro{MSB}{most significant bit}
 \acro{LSB}{least significant bit}
 \acro{PSMC}{partially symmetric monomial code}
\end{acronym}

\begin{abstract}
The automorphism group of a code is the set of permutations of the codeword symbols that map the whole code onto itself. For polar codes, only a part of the automorphism group was known, namely the \ac{LTA}, which is solely based upon the partial order of the code's synthetic channels. Depending on the design, however, polar codes can have a richer set of automorphisms. In this paper, we extend the \ac{LTA} to a larger subgroup of the \ac{GA}, namely the \ac{BLTA} and show that it is contained in the automorphism group of polar codes. Furthermore, we provide a low complexity algorithm for finding this group for a given information/frozen set and determining its size. Most importantly, we apply these findings in automorphism-based decoding of polar codes and report a comparable error-rate performance to that of \ac{SCL} decoding with significantly lower complexity.
\end{abstract}
\acresetall

\section{Introduction} \label{sec:intro}

Polar codes are the first channel codes which are theoretically proven to asymptotically achieve the channel capacity under \ac{SC} decoding \cite{ArikanMain}.
In the short length regime, \ac{CRC}-aided polar codes under \ac{SCL} decoding \cite{talvardyList} achieves an outstanding performance and, thus, selected as the channel code for the uplink and downlink control channel of the 5G standard \cite{polar5G2018}.
Due to the highly symmetric structure of the polar code factor graph, decoders using the concept of factor graph permutations are proposed in \cite{PermutedSCL}, \cite{elkelesh2018belief} and \cite{Moscow_Huawei_Polar_paper}.

A different approach is to use the symmetries in the code itself, i.e., its automorphism group. To this end, polar codes are viewed as decreasing monomial codes \cite{bardet_polar_automorphism}. In \cite{bardet_polar_automorphism}, it is shown that the automorphism group of decreasing monomial codes (and, thus, polar codes) is at least the \ac{LTA}, solely based on a partial order of synthetic channels. This proved to be sufficient for the application of the minimum-weight codeword enumeration. However, in general, we expect decreasing monomial codes to have more automorphisms. This is easily verified by the fact that \ac{RM} codes can be seen as a special case of decreasing monomial codes with an automorphism group known to be the \ac{GA} \cite{macwilliams77}, which is much larger than \ac{LTA}.

Automorphism-based decoding has been successfully applied to \ac{RM} codes \cite{StolteRekursivPlotkin, rm_automorphism_ensemble_decoding} and \ac{BCH} codes \cite{HUBER_BCH_Golay}. However, it was not yet possible to use the automorphism group in \ac{SC}-based decoding of polar codes. The reason for this is that \ac{LTA}-based automorphisms cannot result in any gains under \ac{SC}-based (ensemble) decoding, as proven in \cite[Theorem 2]{rm_automorphism_ensemble_decoding}. Therefore, it is crucial to find automorphisms outside the \ac{LTA} to enable efficient parallel ensemble decoding of polar codes. Further potential applications include analysis of some post-quantum cryptography schemes \cite{bardet_crypto}.

The main contribution of this work is the introduction of a larger automorphism group of decreasing monomial codes, namely the \ac{BLTA}. We provide efficient algorithms for finding this group and sampling from it. The concept applies to polar codes, \ac{RM} codes and the recently proposed \acp{PSMC} \cite{urbanke_monomial_codes}.

\section{Preliminaries} \label{sec:prelim}

\subsection{Polar Codes}
Polar codes are constructed based on the concept of channel polarization \cite{ArikanMain}. $ N $ identical \acp{DMC} are converted, via the channel transform, into $ N $ synthetic channels that show a polarization behavior. This means that a fraction of the bit-channels become very reliable (i.e., noiseless), while the rest of the synthetic bit-channels become totally noisy. Information is transmitted only on the $ K $ most reliable channels (\emph{information channels}), while the poor channels are set to ``0'' (\emph{frozen channels}). This is equivalent to selecting $ K $ rows from the Hadamard matrix $ \mathbf{G}_N = \left[ \begin{smallmatrix}1 & 0 \\ 1 & 1 \end{smallmatrix}\right]^{\otimes n} $ with $ N = 2^n $ to form the generator matrix $\mathbf{G}$ of the code. 

Alternatively, polar codes can be viewed as monomial codes \cite{bardet_polar_automorphism}. In this perspective, each synthetic channel corresponds to a monomial in $ n $ binary variables $ x_i $. The set of all monomials in $ n $ variables is defined as $ \mathcal{M}_n $ and a polar code is a specific subset $ I $, called the \emph{information set} of the polar code. Every monomial can be written as
\ifdef{\negvspaces}{\vspace{-.2cm}}{}
\begin{equation}\label{eq:monomial}
f = \prod_{i \in \ind(f)} x_i \ifdef{\negvspaces}{\vspace{-.2cm}}{}
\end{equation}
where $ \ind(f) $ is an ordered subset of the variable indices $ \Omega = [0, n-1] \triangleq \{0,1, \dots,n-1\} $ and directly corresponds to the $ \ell $-th row of the generator matrix as
\ifdef{\negvspaces}{\vspace{-.2cm}}{}
\begin{equation}\label{eq:row}
 \ell = \sum_{i \in \Omega \setminus \ind(f)} 2^{i}. \ifdef{\negvspaces}{\vspace{-.2cm}}{}
\end{equation}
In other words, the monomial $ f $ corresponds to the row whose binary representation has zeros exactly in the bit-positions of the variables contained in $ f $. A message is a polynomial 
\ifdef{\negvspaces}{\vspace{-.2cm}}{}
\begin{equation}\label{eq:polynomial}
u(x_0, \dots, x_{n-1}) = \sum_{f \in I} u_f \cdot f(x_0, \dots, x_{n-1}) \ifdef{\negvspaces}{\vspace{-.2cm}}{}
\end{equation}
with $ K $ coefficients $ u_f \in \FF_2 $. The respective codeword is given by the evaluation of $ u(\xv) $ in all $ N $ points $ \xv \in \FF_2^n $. As a convention, we assume the $ j $-th codeword symbol is obtained from the point $ \xv $ equal to the binary expansion of $ j $.

\subsection{Partial Order}
It was shown in \cite{bardet_polar_automorphism} and \cite{partialorder} that the synthetic channels exhibit a partial order ``$ \preccurlyeq $'' with respect to their reliability, i.e, $ f \preccurlyeq g $ means that the synthetic channel corresponding to monomial $ f $ is more reliable than the one corresponding to $ g $. For monomials of equal degree this partial order is defined as
\ifdef{\extended}{\ifdef{\negvspaces}{\vspace{-.2cm}}{}}{\ifdef{\negvspaces}{\vspace{-.6cm}}{}}
\begin{equation}\label{def:partial_order1}
f \preccurlyeq g \Leftrightarrow \ind(f)_j \leq \ind(g)_j \quad \forall j = 0, \dots, \deg(f)-1
\end{equation}
and for monomials of different degree
\ifdef{\negvspaces}{\vspace{-.2cm}}{}
\begin{equation}\label{def:partial_order2}
f \preccurlyeq g \Leftrightarrow \exists g^* | g \text{ with } \operatorname{deg}(g^*) = \operatorname{deg}(f) \text{ and } f \preccurlyeq g^*.
\end{equation}

\subsection{Decreasing Monomial Codes}
A \emph{decreasing monomial code} is a polar code whose monomial selection obeys the partial order \cite{bardet_polar_automorphism}. More precisely, if a synthetic channel is selected as an \emph{information channel}, all stronger channels w.r.t. ``$\preccurlyeq $'' are also information channels. Mathematically, this can be written as
\ifdef{\negvspaces}{\vspace{-.25cm}}{}
\begin{equation}\label{def:dec_mon_code}
\forall g \in I, \forall f \in \mathcal{M}_n \text{ with } f \preccurlyeq g \Rightarrow f\in I.\ifdef{\negvspaces}{\vspace{-.2cm}}{}
\end{equation}
Almost all practical polar code constructions result in decreasing monomial codes. A decreasing monomial code can be fully specified by a minimal information set $ I_\mathrm{min} $ containing only a small number of monomials called \emph{generators}. All other monomials are implied by the partial order:
\ifdef{\negvspaces}{\vspace{-.15cm}}{}
\begin{equation}
	I = \bigcup_{g \in I_\mathrm{min}} \left\{f \in \mathcal{M}_n, f\preccurlyeq g\right\}.\ifdef{\negvspaces}{\vspace{-.2cm}}{}
\end{equation}
Moreover, the \ac{RM} code of order $ r $ and length $N=2^n$ (i.e., RM$\left(r,n\right)$-code) is a special case of a decreasing monomial code with $ I_\mathrm{min} = \left\{x_{n-r} \cdots x_{n-1}\right\}$. In this paper, we will notate $ I_\mathrm{min} $ as numerical row indices, according to Eq. (\ref{eq:row}).
\ifdef{\negvspaces}{\vspace{-.1cm}}{}
\subsection{Automorphisms of Decreasing Monomial Codes}
The automorphism group $ \operatorname{Aut}(\mathcal{C}) $ of a code $\mathcal{C}$ is the group of codeword symbol permutations, that leave the code unchanged, i.e., map each codeword onto a codeword that is not necessarily different. It was shown in \cite{bardet_polar_automorphism} that the automorphism group of a decreasing monomial code contains at least $ \operatorname{LTA}(2,n) $, i.e., affine transformations of the variables $ x_i $ in the form $ \xv' = \Am \xv + \bv $, with $ \Am\in \FF_2^{n\times n}$ being a lower triangular matrix with a unit diagonal and arbitrary $ \bv\in \FF_2^n $. 

\ifdef{\negvspaces}{\vspace{-.1cm}}{}
\section{Stabilizers of the Monomial Set}
It was shown in \cite{Doan_2018_Permuted_BP} that the stage-shuffling of the polar factor graph corresponds to a bit-index permutation of both the codeword vector $ \cv $ and the message vector $ \uv $ (including the frozen bits). When viewing such permutations from a monomial code perspective, they exactly correspond to permuting the variables $ x_i $ of the monomials from $ \mathcal{M}_n $. Depending on the polar code construction (i.e., information/frozen set), there may exist permutations that keep the information set $ I $ unchanged, i.e., they \emph{stabilize} it. Such a permutation is directly related to that automorphism of the code, where $ \Am $ in $ \xv' = \Am \xv + \bv $ is the corresponding permutation matrix. 

\textbf{Definition (Stabilizer):}
Let $ S(\Omega) $ be the set of all permutations of $ \Omega $. Then a permutation $ \pi \in S(\Omega)$ \emph{stabilizes} a monomial set $ I $, if and only if
\ifdef{\negvspaces}{\vspace{-.2cm}}{}
\begin{equation}\label{eq:stab}
\forall f \in I \implies f' = \pi(f) \triangleq \prod_{i\in \ind(f)}{x_{\pi(i)}} \in I. \ifdef{\negvspaces}{\vspace{-.2cm}}{}
\end{equation}
In other words, $ I $ remains unchanged when permuting the variable indices in the monomials according to $ \pi $.
Furthermore, let $ \Stab(I) $ denote the set of all permutations with this property. Note that $ \Stab(I) $ is a subgroup of $ S(\Omega) $. We call the stabilizer \emph{trivial}, if it only contains the identity permutation.

In the following, we seek to find $ \Stab(I) $ for a given $ I $ and derive some useful properties.

\textbf{Definition (Minimum and Maximum of a Permutation)}:
Let $ \pi \in S(\Omega) $ be some permutation. The minimum $ \min(\pi) $ and maximum $ \max(\pi) $ are defined by the smallest and largest element not fixed by $ \pi $, i.e.,\ifdef{\negvspaces}{\vspace{-.2cm}}{}
\begin{align}\label{eq:minmax}
\min(\pi) &\triangleq \min \left\{ i \; \middle| \; i \in \Omega, \pi(i)\ne i \right\},\\
\max(\pi) &\triangleq \max \left\{ i \; \middle| \; i \in \Omega, \pi(i)\ne i \right\}.
\end{align}

\textbf{Definition (Interval Disjoint and Interlocked Permutations):}
Two permutations $ \pi_1 $ and $ \pi_2 $ are \emph{interval disjoint}, if the intervals $ \Omega_{\pi_1} = [\min(\pi_1), \max(\pi_1)] $ and $ \Omega_{\pi_2} = [\min(\pi_2), \max(\pi_2)] $ are disjoint. Note that there may be elements of $ \Omega_{\pi_i} $ which are not affected by $ \pi_i $. Permutations are said to be \emph{interlocked}, if they are not interval disjoint and do not share elements.

Let $ C(\pi) $ be the cycle decomposition of $ \pi $. By merging all interlocked cycles $ \sigma \in C(\pi) $ into the same sub-permutations $ \rho_i $, we obtain the \emph{interval disjoint decomposition} $ T(\pi) = \left\{\rho_0, \dots, \rho_{d-1} \right\}, $ as the unique set of pairwise interval disjoint permutations $ \rho_i $ such that $  \pi = \rho_0 \circ \cdots \circ \rho_{d-1} $.

\textbf{Theorem 1 (Stabilizers):}
Let $ I $ be the monomial set of a decreasing monomial code with a non-trivial stabilizer $ \Stab(I) $. Then the following statement holds:

If a non-trivial permutation $\pi$ stabilizes $ I $, then all permutations of the disjoint intervals of $ \pi $ stabilizes $I$ as well, i.e.,
\ifdef{\extended}{\ifdef{\negvspaces}{\vspace{-.2cm}}{}}{\ifdef{\negvspaces}{\vspace{-.5cm}}{}}
\begin{equation}\label{eq:stabilizer_thm}
\pi \in \Stab(I) \Rightarrow \left\langle S\left(\left[\min(\rho),\max(\rho)\right]\right)  \right\rangle_{\rho \in T(\pi)} \subseteq \Stab(I),
\end{equation}
where $ \langle \cdot \rangle $ denotes the \emph{join} of subgroups.

\textit{Proof:} \ifdef{\extended}{The proof is given in Appendix \ref{appx:proof_thm1}.}{The proof is given in the extended version \cite{extended_arxiv}.}

Theorem 1 has a useful corollary revealing the structure of $ \Stab(I) $.

\textbf{Corollary:}
$ \Stab(I) $ can be written as the join of permutations groups $ S(\Omega_k) $ of partitions of $ \Omega $, i.e.,
\ifdef{\negvspaces}{\vspace{-.2cm}}{}
\begin{align}\label{eq:stab_corollary}
 \Stab(I) &= \left\langle S\left(\Omega_0\right), \dots, S\left(\Omega_{m-1}\right)\right\rangle \nonumber\\
 \text{ with }  \bigcup_{k=0}^{m-1}\Omega_k &= \Omega \; \text{ and } \; \Omega_k \cap \Omega_l = \emptyset \text{ for } k\ne l. \vspace{-.2cm}
\end{align}
In other words, every permutation $ \pi \in \Stab(I) $ can be written as a product of (potentially trivial) permutations of the intervals $ \Omega_k $ and vice versa. Note that $ \Omega_k $ may contain only a single element when $ S(\Omega_k) $ does not contribute to any non-trivial permutation.

\textit{Proof:}
Assume the sub-intervals are not disjoint, i.e., there exist two sub-intervals $ \Omega_k $ and $ \Omega_l $ with $ S(\Omega_k) \subseteq \Stab(I) $ and $ S(\Omega_l) \subseteq \Stab(I) $ but $ \Omega_k \cap \Omega_l \ne \emptyset $ and neither $ \Omega_k \subseteq \Omega_l $ nor $ \Omega_l \subseteq \Omega_k $. Then one can pick two permutations (e.g., extremal transpositions) $ \pi_1 \in S(\Omega_k) $ and $ \pi_2 \in S(\Omega_l) $ which are not interval disjoint and $ \pi = \pi_1 \circ \pi_2 $ is either a single cycle or the product of interlocked cycles. In both cases, $ \pi $ stabilizes $ I $ and, thus, $ S(\Omega_k \cup \Omega_l) \subseteq \Stab(I) $. Therefore, every permutation in $ \Stab(I) $ either falls into an existing sub-interval or expands or merges sub-intervals, keeping the partition property. $ \qed $

The partition (and therefore $ \Stab(I) $) is fully described by the list of interval sizes $ \sv = [s_k] $ of the $ m $ sub-intervals $ \Omega_k $, i.e.,
\ifdef{\negvspaces}{\vspace{-.4cm}}{}
\begin{equation}\label{eq:sk}
s_k = |\Omega_k| = \max(\Omega_k) - \min(\Omega_k) + 1.\ifdef{\negvspaces}{\vspace{-.3cm}}{}
\end{equation}

The corollary gives us an algorithm for finding the sub-intervals $ \Omega_k $ for an arbitrary decreasing monomial code with information set $ I $. We know that all permutations in $ S(\Omega_k) $ are contained in $ \Stab(I) $, as we can pick trivial permutations for the other sub-intervals. In particular, also the transposition $ \pi = (\min(\Omega_k),\max(\Omega_k)) $ stabilizes $ I $. Therefore, we can find the borders of the sub-intervals by \emph{systematically} searching for pairs $ i_0, i_1 $ with maximal distance. Algorithm \ref{alg:stab} provides a pseudo-code for this procedure. The algorithm has a worst case runtime of $ \mathcal{O}(K \cdot n^2) $, with the check $ \pi(I)=I $ requiring $ K $ comparisons.
\ifdef{\negvspaces}{\vspace{-.2cm}}{}
\begin{algorithm}[tbh]
	\SetAlgoLined
	\LinesNumbered
	\SetKwInOut{Input}{Input}\SetKwInOut{Output}{Output}
	\Input{Information set $ I $ of decreasing monomial code in $ n $ variables}
	\Output{List of sub-interval sizes $ \sv $}
	$ \sv \leftarrow [\,],\quad i_0 \leftarrow 0 $\;
	\While{$ i_0 < n $}{
		$ i_1 \leftarrow n-1 $\;
		\While{$ i_1 \ge i_0 $}{
			$ \pi \leftarrow (i_0,i_1) $\;
			\uIf{$\pi(I) = I$}{
				append $ i_1 - i_0 + 1 $ to $ \sv $\;
				$ i_0 \leftarrow i_1 + 1 $\;
			}
			\Else{
				$ i_1 \leftarrow i_1 - 1 $\;
			}
		}
	}
	\caption{\footnotesize Finding $ \Stab(I) $ in terms of the partition of $ \Omega $ into sub-intervals of size $ s_k $.}\label{alg:stab}
\end{algorithm}

We can represent $ \Stab(I) $ as a set of $ n \times n $ permutation matrices $ P_\sv(n) $, where $ \sv = [s_k] $ defines a block diagonal structure with blocks of sizes $ s_k \times s_k $. Except for the block diagonal elements, all matrix elements are zero. For a non-trivial stabilizer, we hereby find automorphisms \emph{outside} \ac{LTA}, as no permutation matrix is lower-triangular besides the identity permutation. 

\section{The Automorphism Group of Polar Codes}
In the following, we combine both \ac{LTA} and the newly found stabilizer group into a larger group, namely the \emph{block lower-triangular affine group} (BLTA).

\textbf{Definition (Block Indices):}
We denote a partition of the interval $ \Omega = [0,n-1] $ by a sequence of $ m $ positive integers $ s_k > 0 $ for the sizes of the sub-intervals. The \emph{interval start} $ \gamma_k $ is the first element of the $ k$-th sub-interval and is defined as the cumulative sum
\ifdef{\negvspaces}{\vspace{-.2cm}}{}
\begin{equation}\label{gamma}
\gamma_k = \sum_{i=0}^{k-1} s_k.\ifdef{\negvspaces}{\vspace{-.2cm}}{}
\end{equation}

The \emph{index function} $ k(i) $ returns the index of the sub-interval that contains $ i $ and is defined as $ k(i) = \max\left\{k: \; i  \ge \gamma_k \right\} $.

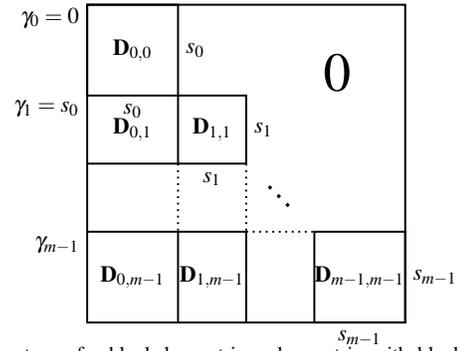
\begin{figure}[tb]
	\centering
	\resizebox{.7\columnwidth}{!}{\input{tikz/blt.tikz}}
	\ifdef{\negvspaces}{\vspace{-.4cm}}{}
	\caption{\footnotesize Structure of a block lower-triangular matrix with block sizes $ s_k $ and \emph{block starts} $ \gamma_k $.}
	\label{fig:BLT_Structure}
	\ifdef{\negvspaces}{\vspace{-.4cm}}{}
\end{figure}

\textbf{Definition (Block Lower-Triangular Matrix):} An $ n \times n $ matrix $ \mathbf{A} $ over an arbitrary field is \emph{block lower-triangular} with block sizes $ \sv = [s_k], 0 \le k \le m-1 $, if all elements to the right of the block diagonal are zero, i.e., $ a_{i,j} = 0 \; \forall j \ge \gamma_{k(i)}+ s_{k(i)} $. 

The blocks of the matrix are denoted by $ \Dm_{k,l} $. Fig. \ref{fig:BLT_Structure} shows the general structure of a block lower diagonal matrix. As square block matrices naturally extend conventional matrices, we have the following properties:

\begin{enumerate}
	\item The product of two block lower-triangular matrices is also a block lower-triangular matrix with the same block structure.
	\item A block lower-triangular is non-singular if and only if all blocks on the main diagonal $ \Dm_{k,k} $ are non-singular. 
	\item The inverse of a block lower-triangular matrix is also a block lower-triangular matrix with the same block structure as the original matrix.
\end{enumerate}

As a consequence, non-singular block lower-triangular matrices form a group under matrix multiplication. Note that associativity is inherited from matrix multiplication and the identity matrix $ \mathbf{I} $ is always block lower-triangular. The size of this group can be easily computed in terms of $ \sv $. For this, observe that in row $ i $, there are $ \gamma_{k(i)}+s_{k(i)} $ elements that can be 0 or 1 each. However, one has to deduce the number of cases where the row is a linear combination of the $ i $ previous rows. Therefore, the number of invertible block lower-triangular matrices is
\ifdef{\negvspaces}{\vspace{-.3cm}}{}
\begin{align}
	N_\mathrm{IBLT}(\sv) &= \prod_{i=0}^{n-1}\left(2^{\gamma_{k(i)}+s_{k(i)}} - 2^i\right) \label{eq:num_whole_perspective}\\
	&= \prod_{k=0}^{m-1}\prod_{i'=0}^{s_k-1}\left(2^{\gamma_k+s_k} - 2^{\gamma_k+i'}\right) \nonumber\\
	&= \prod_{k=0}^{m-1}\left( 2^{\gamma_k \cdot s_k} \prod_{i'=0}^{s_k-1}\left(2^{s_k} - 2^{i'}\right) \right).\label{eq:num_block_perspective}
\end{align}
While Eq. (\ref{eq:num_whole_perspective}) expresses the number from a whole matrix perspective, Eq. (\ref{eq:num_block_perspective}) views the same thing from a block matrix perspective. In particular, the inner product gives the number of non-singular diagonal blocks $ \Dm_{k,k} $, while $ 2^{\gamma_k \cdot s_k} $ is the number of \emph{arbitrary} rectangular matrices to the left of the block diagonal $ \Dm_{k,l} $ with $ l<k $, for each block row $ k $.

\textbf{Definition (Block Lower-Triangular Affine Group, BLTA):}
The block lower-triangular affine group $ \operatorname{BLTA}(\sv, n) $ is the set of affine transformations $ \xv' = \Am \xv + \bv $ over $ \FF_2^n $ with $ \Am \in \FF_2^{n\times n}$ non-singular block lower-triangular with block structure $ \sv $ and an arbitrary $ \bv \in \FF_2^n $.

From the discussion of block lower-triangular matrices above,  it is easy to see that \ac{BLTA} is indeed a group, in particular a subgroup of $ \operatorname{GA}(2,n) $. Moreover, it can be seen that $ \operatorname{LTA}(2,n) $ and $ \operatorname{GA}(2,n) $ are themselves special cases of $ \operatorname{BLTA}(\sv,n) $, with $ \sv = [1,\dots,1] $ and $ \sv = [n] $, respectively. 

\textbf{Lemma 1:}
The join of the group of block-permutation transformations $ P_\sv(n) $ and the group lower triangular affine transformations $ \operatorname{LTA}(2,n) $ is exactly $ \operatorname{BLTA}(\sv,n) $, i.e.,
\ifdef{\negvspaces}{\vspace{-.1cm}}{}
\begin{equation}\label{eq:lemma_join}
\operatorname{BLTA}(\sv,n) = \langle P_\sv(n), \operatorname{LTA}(2,n)  \rangle.\ifdef{\negvspaces}{\vspace{-.1cm}}{}
\end{equation}
In other words, any composition of transformations from $ \operatorname{LTA}(2,n) $ and $ P_\sv(n) $ is a transformation from $ \operatorname{BLTA}(\sv,n) $ and vice versa.

\textit{Proof:} ``$ \Rightarrow $'': Obviously, $ P_\sv(n) \subseteq \operatorname{BLTA}(\sv,n) $, as permutation matrices are non-singular and the block structure is given. Similarly, $ \operatorname{LTA}(2,n) \subseteq \operatorname{BLTA}(\sv,n) $, where again we have a special case of affine transformations. Also, as $ \operatorname{BLTA}(\sv,n) $ is closed, a composition of any transformations will not generate any elements outside $ \operatorname{BLTA}(\sv,n) $.

``$ \Leftarrow $'': We can show this by observing that any block lower-triangular matrix $ \Am $ may be decomposed as $ {\Am = \Pm_1\cdot \Lm_1 \cdot \Pm_2 \cdot \Lm_2 \cdot \Pm_3} $, with $ \Pm_i \in P_\sv(n) $ and $ \Lm_i \in \operatorname{LTA}(2,n) $. For this, consider the LUP decomposition of $ \Am $, i.e., $ \Pm \Am = \Lm \Um $ \cite{cormen_algorithms_2001}. The block lower-triangular structure of $ \Am $ ensures that also $ \Um $ and $ \Pm $ are \emph{block} lower-triangular. One can now transform $ \Um $ into a conventional lower-triangular matrix by reversing the order of the rows and columns within each block. This can be written as $ \Lm_2 = \Pm_{\mathrm{BR}}(\sv) \cdot \Um \cdot \Pm_{\mathrm{BR}}(\sv) $, with $ \Pm_{\mathrm{BR}}(\sv) = [p_{i,j}] $ and
\ifdef{\negvspaces}{\vspace{-.2cm}}{}
\begin{equation}\label{eq:pbr}
p_{i,j} = \begin{cases}
1 & \text{for } j = 2\gamma_{k(i)}+s_{k(i)} - 1 - i\\
0 & \text{else}
\end{cases}.
\end{equation}
Finally, as $ \Pm_{\mathrm{BR}}(\sv) = \Pm_{\mathrm{BR}}^{-1}(\sv) $, we have $ \Pm_1 = \Pm^{-1} $, $ \Pm_2 = \Pm_3 = \Pm_{\mathrm{BR}}(\sv) $ and $ \Lm_1 = \Lm $. The additive term $ \bv $ may be included (i.e., also properly permuted) in any of the \ac{LTA} transformations. $ \qed $

\textbf{Theorem 2 (Automorphisms of Polar Codes):}
Let $ \mathcal{C} $ be a decreasing monomial code in $ n $ variables with information set $ I $. Then 
\begin{equation}\label{eq:blta_is_aut}
 \operatorname{BLTA}(\sv,n) \subseteq \operatorname{Aut}(\mathcal{C}) 
\end{equation} with $ \sv $ being the block structure of $ \Stab(I) $.

\textit{Proof:} The proof directly follows from Lemma 1, as both \ac{LTA} and $ \Stab(I) $ correspond to automorphisms of the code. $ \qed $

We furthermore conjecture that Eq.~(\ref{eq:blta_is_aut}) holds with equality when considering only affine automorphisms. However, we were not yet able to find a rigorous proof. To prove it, one would need to show that for every nonzero element $ a_{i,j} $ with $ i<j $ in an affine transformation $ (\Am, \bv) \in \operatorname{Aut}(\mathcal{C}) $, the variable permutation $ \pi = (i,j) $ must be contained in $ \Stab(I) $.\footnote{This conjecture has been proven in \cite{li2021_proof_of_conjecture} while this paper was under review.}

\subsection{Number of Automorphisms}
The number of automorphisms is (at least) the size of $ \operatorname{BLTA}(\sv,n) $ for a code with block structure $ \sv $. Clearly, this is the number of non-singular block lower-triangular matrices times the number of affine translations $ \bv $. Using Eq. (\ref{eq:num_block_perspective}), we have
\ifdef{\negvspaces}{\vspace{-.2cm}}{}
\begin{equation}
|\operatorname{BLTA}(\sv,n)| = N_\mathrm{IBLT}(\sv)\cdot 2^{n} = 2^n \cdot \prod_{k=0}^{m-1}\left( 2^{\gamma_k \cdot s_k} \prod_{i'=0}^{s_k-1}\left(2^{s_k} - 2^{i'}\right) \right).
\end{equation}

Note that this equates to the sizes of $ \operatorname{LTA}(2,n) $ and $ \operatorname{GA}(2,n) $ for the special cases $ \sv = [1,\dots,1] $ and $ \sv = [n] $, respectively. 

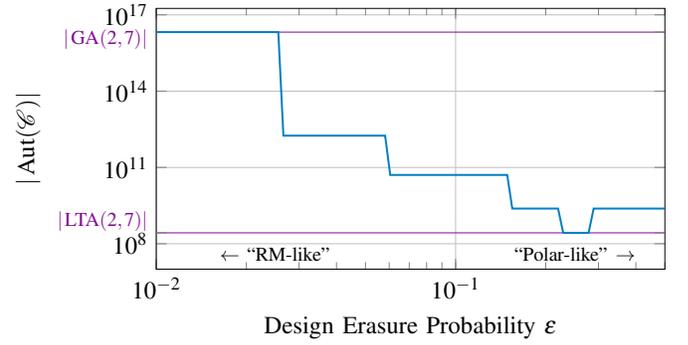
\begin{figure}[tb]
	\centering
	\resizebox{\columnwidth}{!}{\input{tikz/multiplicity_chart_bhattacharyya.tikz}}
	\ifdef{\negvspaces}{\vspace{-.6cm}}{}
	\caption{\footnotesize Number of automorphisms of $ (128,64) $ polar codes with Bhattacharyya-based construction versus \ac{BEC} design erasure probability $ \epsilon $.}
	\label{fig:amc_bhatta}
	\ifdef{\negvspaces}{\vspace{-.4cm}}{}
\end{figure}

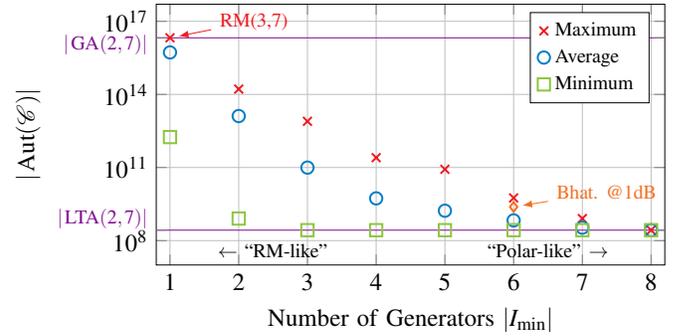
\begin{figure}[tb]
	\centering
	\resizebox{\columnwidth}{!}{\input{tikz/multiplicity_chart_generators.tikz}}
	\ifdef{\negvspaces}{\vspace{-.6cm}}{}
	\caption{\footnotesize Maximum, average and minimum number of automorphisms of \emph{all} $ (128,64) $ decreasing monomial codes versus their number of generators $ |I_\mathrm{min}| $.}
	\label{fig:amc_imin}\ifdef{\negvspaces}{\vspace{-.5cm}}{}
\end{figure}

Fig. \ref{fig:amc_bhatta} shows the sizes of the automorphism groups for polar codes with $ N=128 $ and $ K=64 $ designed according to the Bhattacharyya parameter of the synthetic channels. This construction assumes a \ac{BEC} with erasure probability $ \epsilon $. It can be seen that for low erasure probability, this construction generates the RM(3,7)-code. The larger the values of $ \epsilon $, the fewer the automorphisms featured by the code.
In Fig. \ref{fig:amc_imin}, we evaluate the influence of the number of generators of a code $ |I_\mathrm{min}| $ on the size of the automorphism group, also for the case of $ N=128 $ and $ K=64 $. Since there exist usually many codes with the same number of generators, we plot the minimum, average and maximum automorphism group sizes for each value of $ |I_\mathrm{min}| $. To obtain these numbers, we enumerated all 1007 $ (128,64) $ decreasing monomial codes using a tree search. As just mentioned, it can be seen that a smaller size of $ I_\mathrm{min} $ generally results in a larger number of automorphisms. We find the \ac{RM} code on the very left of the plot, while typical polar codes lie more towards the right edge. It is worth mentioning that, from a code design perspective, several code constructions can be viewed as lying between polar and \ac{RM} codes (e.g., \cite{urbanke_monomial_codes}, \cite{HybridTse}, \cite{RMurbankePolar} and \cite{GenAlg_Journal}).

\vspace{0.5cm}

\subsection{Sampling Automorphisms}
For some practical applications such as automorphism ensemble decoding \cite{rm_automorphism_ensemble_decoding}, it is required to sample from the automorphism group, i.e., to pick a permutation from $ \operatorname{BLTA}(\sv,n) $ at random. In general, it is difficult to ensure that a random matrix is invertible. If the fraction of non-singular matrices out of all matrices is sufficiently large, one can generate random matrices and test for invertibility. For binary matrices (GL for general linear), this probability is lower bounded \cite{OEISA048651} as
\ifdef{\negvspaces}{\vspace{-.2cm}}{}
\begin{align}
	p_{\mathrm{succ,GL}} &=  \frac{\prod_{i=0}^{n-1}\left(2^n-2^i\right)}{2^{(n^2)}} = \prod_{i'=1}^{n}\left(1-2^{-i'}\right) \nonumber \\
	&\ge \lim_{n\to\infty} \prod_{i'=1}^{n}\left(1-2^{-i'}\right) = 0.28878\dots \label{eq:psucc_gl}
\end{align}
However, the same expression for a block lower-triangular (BLT) matrices, i.e.,
\ifdef{\negvspaces}{\vspace{-.2cm}}{}
\begin{align}
p_{\mathrm{succ,BLT}} & = \frac{\prod_{i=0}^{n-1}\left(2^{\gamma_{k(i)}+s_{k(i)}} - 2^i\right) }{\prod_{i=0}^{n-1}2^{\gamma_{k(i)}+s_{k(i)}} } = \prod_{i=0}^{n-1}\left(1-2^{i-\gamma_{k(i)}-s_{k(i)}}\right) \nonumber \\
&\ge \prod_{i=0}^{n-1}\left(1-2^{-1}\right) = 2^{-n} \xrightarrow{n \to \infty} 0, \label{eq:psucc_blt}
\end{align}
cannot be lower bounded, since the last line holds with equality for the case $ \sv = [1,\dots,1] $. We therefore propose a different method, based on the fact that only the blocks on the diagonal must be non-singular:
\begin{enumerate}
\item For $k = 0,1,\cdots, m-1$, sample the square blocks $ \Dm_{k,k} $ on the main diagonal from $ \operatorname{GL}(2,s_k) $, i.e., generate random $ s_k \times s_k $ binary matrices until a non-singular one is found, with success probability $ p_{\mathrm{succ},k} = \prod_{i=1}^{s_k}\left(1-2^{-i}\right) $.
\item Select all elements below the block diagonal (i.e., $ a_{i,j} $ with $ j < \gamma_{k(i)} $, or blocks $ \Dm_{k,l} $ with $ l<k $) randomly uniformly from $ \{0,1\} $. 
\end{enumerate}
This method has the advantage that each block on the diagonal can be independently sampled, resulting in total in the same lower bound, Eq.~(\ref{eq:psucc_gl}), which is fulfilled with equality for the worst case of $ \sv = [n] $.
	
\section{Polar Codes under Automorphism SC Decoding}

\begin{figure}
	\centering
	\resizebox{0.975\columnwidth}{!}{\input{tikz/polar_256_aut_sc.tikz}}
	\ifdef{\negvspaces}{\vspace{-.4cm}}{}
	\caption{\footnotesize Comparison of $\left(N=256,K=128\right)$ polar codes under SC, Aut-SC and SCL decoding; BI-AWGN channel. \ifdef{\extended}{Appendix \ref{appx:128_results} gives \ac{BLER} results for (128,64) polar codes.}{In the extended version \cite{extended_arxiv}, we provide more \ac{BLER} results for the (128,64) polar codes.}}
	\label{fig:Polar256_AutSC_BLER}\ifdef{\negvspaces}{\vspace{-.5cm}}{}
\end{figure}
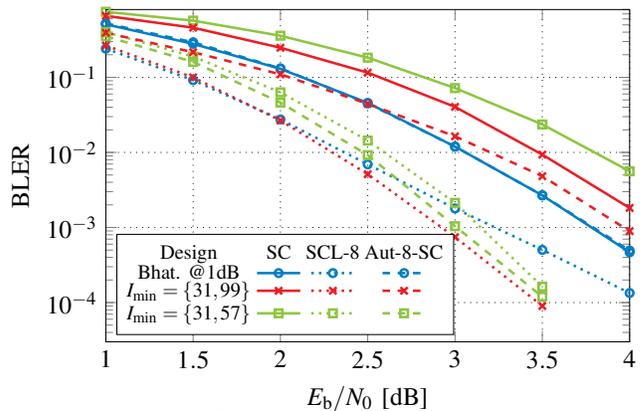

\begin{table}
	\centering\begin{tabular}{c|cc|cc}
		Design & $ \sv $ & $ |\operatorname{Aut}(\mathcal{C})| $& $ d_\mathrm{min} $\textsuperscript{a} & $ A_{d_\mathrm{min}} $\textsuperscript{b} \\
		\hline
		Bhat. @1 dB & $[2, 1, 1, 1, 1, 1, 1]$ & $2.06\cdot10^{11}$ & 8 & 96 \\
		$I_\mathrm{min}=\{31,99\}$ &$[5, 3]$ & $1.41\cdot10^{16}$ & 16 & 69936 \\
		$I_\mathrm{min}=\{31,57\}$ &$[3, 5]$ & $1.41\cdot10^{16}$ & 16 & 69936\\
		
	\end{tabular}
	\ifdef{\negvspaces}{\vspace{-.2cm}}{}
		\caption{\footnotesize Properties of the compared (256,128) polar codes. \textsuperscript{a}$ d_\mathrm{min} $: minimum distance of the code. \textsuperscript{b}$ A_{d_\mathrm{min}} $: number of minimum-weight codewords.}
	\label{tab:code_props_256}
	\ifdef{\negvspaces}{\vspace{-.6cm}}{}
\end{table}

As an application, we now evaluate polar codes under automorphism SC (Aut-SC) decoding. As proposed in \cite{rm_automorphism_ensemble_decoding}, we use $ M=8 $ parallel independent \ac{SC} decoders, each decoding a permuted version of the received sequence $ \yv $. The codeword estimates of each \ac{SC} decoder are un-permuted and the ML-in-the-list method is applied to select the final codeword estimate. The permutations are conducted by automorphisms randomly sampled from the \ac{BLTA} group of the particular code, found using Algorithm~\ref{alg:stab}. Note that this decoder is similar to \ac{SCL}, however, \emph{no sorting} of the path-metrics are required, as the constituent decoders are independent. We assume an \ac{AWGN} channel with \ac{BPSK} modulation.

Fig. \ref{fig:Polar256_AutSC_BLER} shows the \ac{BLER} performance of (256,128) polar codes under SC-based decoding. In particular, we compare plain \ac{SC} decoding \cite{ArikanMain} with \ac{SCL} with list size 8 (\ac{SCL}-8) decoding \cite{talvardyList} and Aut-8-SC decoding \cite{rm_automorphism_ensemble_decoding}. First, we see that while being the best code under SC-decoding, the Bhattacharyya construction at design SNR of 1 dB  ($ I_\mathrm{min}=\{59, 79, 105, 149, 163, 224\} $) does not show any gains for Aut-SC decoding, as expected for such few automorphisms outside \ac{LTA}. Next, in order to have a large automorphism group, we designed codes by selecting two generators $ I_\mathrm{min}=\{31,57\} $ and $ I_\mathrm{min}=\{31,99\} $, under the constraint of $ K=128 $. Both codes can be viewed as examples of \acp{PSMC} \cite{urbanke_monomial_codes}. While the \ac{SC} performance degrades, now a significant performance gain is achieved by both Aut-SC and SCL. However, the two constructions show a very different behavior. While for the code with $ I_\mathrm{min}=\{31,99\} $ \ac{SCL} shows a very good performance, Aut-SC shows only small gains. The code with $ I_\mathrm{min}=\{31,57\} $ can, however, outperform \ac{SCL}. Therefore, a strict correlation between \ac{SCL} and Aut-SC decoding performance for partially symmetric codes cannot be inferred and code design for both decoders remains an open problem. 
Table~\ref{tab:code_props_256} lists the parameters and properties of the compared codes.\end{NoHyper}\footnote{An interactive demo of the code properties is provided online: \url{http://webdemo.inue.uni-stuttgart.de/webdemos/08_research/polar/index.php?id=12}}
\begin{NoHyper}
\ifdef{\extended}{In Appendix \ref{appx:128_results} we provide more \ac{BLER} results for the case of $ N=128 $ and $ K=64 $.}{In the extended version of this paper \cite{extended_arxiv}, we provide more \ac{BLER} results for the case of $ N=128 $ and $ K=64 $.} We want to emphasize again that the usage of just \ac{LTA} permutations would result in the \ac{BLER} performance curves of Aut-SC to coincide with plain \ac{SC} decoding as depicted and discussed in \cite{rm_automorphism_ensemble_decoding}. 

\section{Conclusion} \label{sec:conc}
We show that decreasing monomial codes have at least \ac{BLTA} as their automorphism group, which is in most cases larger than the previously known subgroup \ac{LTA}, and propose an algorithm to find this group.
While the automorphisms from \ac{LTA} were proven to yield \emph{no} error-rate performance gains under automorphism-based \ac{SC} decoding when compared to plain \ac{SC} decoding, the newly found \ac{BLTA} permutations show significant gains, outperforming the state-of-the-art \ac{SCL} in some scenarios, with a strictly lower complexity. 

\bibliographystyle{IEEEtran}
\bibliography{references}

\ifdef{\extended}{
\begin{appendix}
\subsection{Proof of Theorem 1}\label{appx:proof_thm1}
Let $ \sigma \in C(\pi) $ be a cycle of the cycle decomposition of $ \pi $ and $ \Omega_\sigma = [\min(\sigma), \max(\sigma)] $ be the interval $ \sigma $ acts on. The index set $ \Omega_\pi $ is the union of the intervals of the cycle decomposition of $ \pi $, i.e.,
\begin{equation}\label{eq:operating interval}
\Omega_\pi = \bigcup_{\sigma \in C(\pi)} \Omega_\sigma,
\end{equation}
and $ \Omega_\pi^c = \Omega \setminus \Omega_\pi $ its complement.
We can factor every monomial $ f\in I $ into a part $ f_\pi $ corresponding to $ \pi $ and a residual:
\begin{equation}
f = \prod_{i\in \ind(f) \cap \Omega_\pi} x_i \prod_{i\in \ind(f) \cap \Omega_\pi^c} x_i = f_\pi \cdot f_{\pi^c}. 
\end{equation}

Now, partition $ I $ into subsets $ I_f $ of the same degree and equal residual $ f_{\pi^c} $, i.e., under the following equivalence relation:
\begin{align}
&f \sim f' \Leftrightarrow \deg(f) = \deg(f') \text{ and } f_{\pi^c} = f_{\pi^c}' \\
&I/{\sim} = \left\{[f]_{\sim} \; \middle| \; f \in I\right\}.
\end{align}
We focus on some monomial $ f $ with subset $ I_f = [f]_{\sim} $. From \cite[Proposition 2]{bardet_polar_automorphism}, we know that within each subset, all elements are comparable under the partial order and it is sufficient to look at the part of $ f $ that is not shared by the elements in $ I_f $, i.e., $ f_\pi $. We will now show that $ I_f $ contains \emph{all} monomials that share $ f_{\pi^c} $ and have the same degrees in intervals of the interval disjoint decomposition of $ \pi $, i.e.,
\begin{equation}\label{eq:whatweshow}
I_f = \left\{g\cdot f_{\pi^c} \in \mathcal{M}_n, \; \deg(g_\rho)=\deg(f_\rho) \; \forall \rho \in T(\pi)\right\},
\end{equation}
by repeatedly applying $ \pi $ and using the partial order.

First, assume $ \pi = \sigma $ is just a single cycle and $ \Omega_\sigma = [i_0,i_1] $. Let $ d=\deg(f_\sigma) $. If $ d=0 $, Eq. (\ref{eq:whatweshow}) is already fulfilled, as $ f $ is the only such monomial. If $ d>0 $, observe:
\begin{enumerate}
	\item After a maximum of $ \ord(\sigma) $ steps, we can transform $ f $ into some $ f' $ with the property $ i_1 \in \ind(f') $.
	\item After a maximum of $ 2d\cdot\ord(\sigma) $ steps, we can transform $ f $ into $ \hat{f} = x_{i_1-d+1} \cdots x_{i_1}$ which is the maximum monomial (w.r.t. the partial order) in $ \Omega_\sigma $.
\end{enumerate}
In both scenarios, a \emph{step} refers to one application of the partial order (i.e., transforming $ f $ into some $ f'\preccurlyeq f $ with $ f' \sim f $) followed by one application of the permutation $ \pi $ (i.e., transforming $ f $ into $ f' = \pi(f) $; $ f'\sim f $ implicitly fulfilled). Both operations will map $ f $ to another monomial $ f' $ that is contained in $ I $, as we assume $ \pi \in \Stab(I) $ and $ I $ belongs to a decreasing monomial code. 

Observations 1) and 2) can be verified by looking at the following algorithm. We denote the monomial at step $ j $ by $ f_j $. Assume, we know some upper limit of $ Q $ steps, in which we can arrive certainly at the desired target state $ \hat{f}=f_Q $. At each step $ j $, find the positions $ \ind(f_j) $ that are not correct if $ \pi $ is applied another $ Q-j $ times, i.e.,
\begin{equation}\label{eq:fixset}
F_j = \left\{ i \in \ind(\pi^{Q-j}(\hat{f})), i\notin \ind(f_j) \right\}.
\end{equation}
After each set of two revolutions of $ \pi $ (or $ 2\cdot\ord(\pi) $ steps), we can remove (at least) one element from $ F_j $, since, as long as we have not yet arrived at the target state, there is an ``empty place'' in the positions affected by $ \pi $ which is permuted to $ i_0 $ at some point. The partial order allows us to move one of the $ x_i $ that are not yet in the target position, into $ i_0 $, as $ x_{i_0} \preccurlyeq x_i $. After another maximum of $ \ord(\pi)-1 $ steps, the respective $ x_i $ has moved to position $ i_1 $, and can be placed in the target position $ \hat{i} $ by the partial order, again, because  $ x_{\hat{i}} \preccurlyeq x_{i_1} $. Therefore, no more than $ Q=2d \cdot \ord(\pi) $ steps are required. Note that this is a very loose upper bound, but to prove Eq. (\ref{eq:whatweshow}), we only need the algorithm to be deterministic and stop after a finite number of steps.

To summarize, we can move any variable in $ f $ to any other place in $ \Omega_\pi $, as we can move it to $ i_0 $ using the partial order, rotate it to $ i_1 $ using the cycle permutation $ \pi $, and then place it in the target position as all indices are reachable by the partial order from $ i_1 $. Due to the resemblance of the cycle $ \pi $ with an irregular paternoster elevator running around a building with floors $ \Omega_\pi $, we call this algorithm the \emph{crazy paternoster algorithm}.

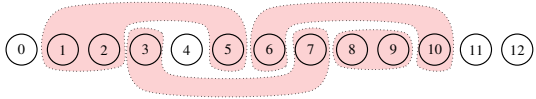
\begin{figure}[tb]
	\centering
	\resizebox{.8\columnwidth}{!}{\input{tikz/interlocked_cycles.tikz}}
	\caption{\footnotesize Visualization of interlocked cycles in the example permutation $ {\pi = (1,5,2)(3,7)(6,10)(8,9)} $.}
	\label{fig:interlocked_cycles}
\end{figure}

The described method can be extended to the case where $ \pi $ is a product of interlocked cycles. We can classify all cycles as either \emph{transit cycles} or \emph{parking cycles}. A parking cycle $ \sigma $ is fully enclosed by another cycle, i.e., $ \exists \sigma' \in C(\pi) $ with $ \Omega_{\sigma} \subset \Omega_{\sigma'} $; while transit cycles partially overlap with another. It is easy to see that there exists a chain of transit cycles $ \sigma_1, \cdots, \sigma_t $ with the properties $ \min(\sigma_1) = i_0 $, $ \max(\sigma_j) > \min(\sigma_{j+1}) $ and $ \max(\sigma_t)=i_1 $. Fig.~\ref{fig:interlocked_cycles} shows the interlocked cycles for the example of $ {\pi = (1,5,2)(3,7)(6,10)(8,9)} $. Here, the cycles $ \sigma_1=(1,5,2) $, $ \sigma_2=(3,7) $ and $ \sigma_3=(6,10) $ form the chain of transit cycles, while $ (8,9) $ fully overlaps with $ (6,10) $ and therefore is classified as a parking cycle.
Using this chain of transit cycles, we can again move any variable in $ f $ to any other position within $ \Omega_\pi $, as the overlap of the cycles allows $ x_i $ to ``change'' from one cycle to the next higher cycle. Note that depending on the degrees of monomials, the order in which the $ x_i $ are moved, must be adjusted. The maximum number of steps remains upper bounded, however, by the same number, namely
\begin{equation}\label{eq:q_ext1}
Q' = 2d\cdot |C(\pi)| \cdot \max_{\sigma \in C(\pi)} \left\{\ord(\sigma)\right\}.
\end{equation}
For the most general case, i.e., if $ \pi $ is a product of multiple interval disjoint permutations, we can use the upper bound
\begin{equation}\label{eq:q_ext2}
Q'' = \max_{\rho \in T(\pi)} \left\{ 2d\cdot |C(\rho)| \cdot \max_{\sigma \in C(\rho)} \left\{\ord(\sigma)\right\}\right\},
\end{equation}
as each interval disjoint region $ \Omega_\rho = [i_{\rho,0}, i_{\rho,1}] $ can be optimized independently according to the procedure above. For all regions, the same backtracking $ \pi^{Q''-j}(\hat{f}) $ is used in each step $ j $. Therefore, Eq. (\ref{eq:whatweshow}) holds for all cases of monomials $ f $. This means that \emph{all} permutations of the intervals $ \Omega_\rho $ stabilize $ I_f $. As this holds for all $ I_f $ individually, it also holds for their union $ I $. $ \qed $

\subsection{Error-Rate Performance for (128,64) Codes}\label{appx:128_results}
\begin{figure}
	\centering
	\resizebox{0.975\columnwidth}{!}{\input{tikz/polar_128_aut_sc.tikz}}
	\caption{\footnotesize Comparison of $\left(N=128,K=64\right)$ polar codes under SC, Aut-SC and SCL decoding; BI-AWGN channel.}
	\label{fig:Polar128_AutSC}
\end{figure}
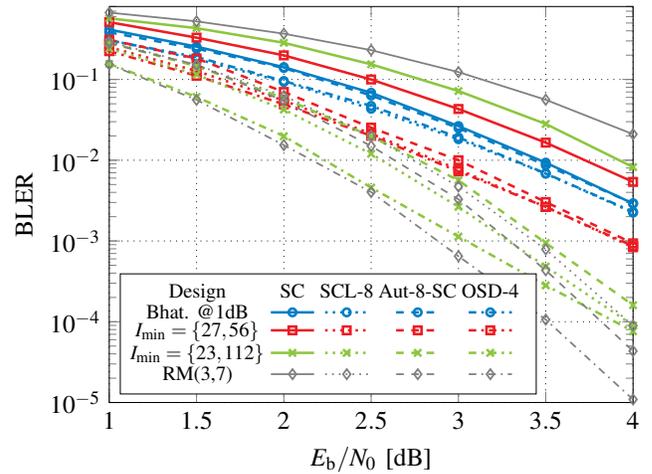

\begin{table}
	\centering\begin{tabular}{c|cc|cc}
		Design & $ \sv $ & $ |\operatorname{Aut}(\mathcal{C})| $& $ d_\mathrm{min} $ & $ A_{d_\mathrm{min}} $ \\
		\hline
		Bhat. @1 dB & $[2, 2, 1, 1, 1]$ & $2.42\cdot10^9$ & 8 & 688 \\
		$I_\mathrm{min}=\{27,56\}$ &$[3, 4]$ & $1.78\cdot10^{12}$ & 8 & 240 \\
		$I_\mathrm{min}=\{23,112\}$ &$[4, 3]$ & $1.78\cdot10^{12}$ & 8 & 16 \\
		RM(3,7) & $[7]$ & $2.10\cdot10^{16}$ & 16 & 94488 \\
		\end{tabular}
	\caption{\footnotesize Properties of the compared (128,64) polar codes.}
	\label{tab:code_props_128}
\end{table}

Fig. \ref{fig:Polar128_AutSC} shows the \ac{BLER} performance of (128,64) polar codes under SC-based decoding. In particular, we compare plain \ac{SC} decoding \cite{ArikanMain} with \ac{SCL} decoding with list size 8 (\ac{SCL}-8) \cite{talvardyList} and Aut-8-SC decoding \cite{rm_automorphism_ensemble_decoding}. Furthermore, \ac{OSD}-4 results serve as an upper bound on the \ac{ML} performance of each code \cite{OSD}. Again, the Bhattacharyya construction at design SNR of 1~dB ($ I_\mathrm{min}=\{31, 45, 51, 71, 84, 97\} $) does not show any gains for Aut-SC decoding. Also, note that the gains of SCL-8 when compared to SC are also smaller than 0.2 dB. Next, we designed codes by selecting two generators $ I_\mathrm{min}=\{27,56\} $ and $ I_\mathrm{min}=\{23,112\} $ in order to have a large automorphism group, under the constraint of $ K=64 $. Both codes can be viewed as examples of \acp{PSMC} \cite{urbanke_monomial_codes}. While the \ac{SC} performance degrades, now a significant performance gain is achieved by both Aut-SC and SCL decoding with comparable performance. In particular, Aut-SC is within 0.1 dB to 0.2 dB of the SCL performance. For completeness, we also included performance results for the \ac{RM} code construction ($ I_\mathrm{min}=\{15\} $). As previously reported in \cite{rm_automorphism_ensemble_decoding}, in the RM case, automorphism-based decoding can even outperform \ac{SCL} decoding. Table~\ref{tab:code_props_128} lists the parameters and properties of the compared (128,64) codes.

\end{appendix}
}{}

\end{NoHyper}

\end{document}

%% file: corporateColours.tex
\definecolor{mittelblau}{RGB}{0, 126, 198}
\definecolor{violettblau}{cmyk}{0.9, 0.6, 0, 0}
\definecolor{rot}{RGB}{238, 28 35}
\definecolor{apfelgruen}{RGB}{140, 198, 62}
\definecolor{gelb}{RGB}{255, 229, 0}
\definecolor{orange}{RGB}{244, 111, 33}
\definecolor{pink}{RGB}{237, 0, 140}
\definecolor{lila}{RGB}{128, 10, 145}
\definecolor{hellgrau}{RGB}{224, 224, 224}
\definecolor{mittelgrau}{RGB}{128, 128, 128}
\definecolor{dunkelgrau}{RGB}{80,80,80}
\definecolor{anthrazit}{RGB}{19, 31, 31}
\definecolor{darkgreen}{RGB}{34,139,34}

%% file: nodesAndStyles.tex
\tikzset{
       vnd/.style={
        shape=circle,
        fill=black,
        draw,
        inner sep=0pt,
        minimum size=0.2cm},
        cnd/.style={
        shape=rectangle,
        fill=white,
        draw,
        minimum width=0.05mm,
        minimum height = 0.05mm}, 
         vndR/.style={
        shape=circle,
        fill=red,
        draw,
        inner sep=0pt,
        minimum size=0.2cm},
        cndR/.style={
        shape=rectangle,
        fill=white,
        draw=red,
        minimum width=0.05mm,
        minimum height = 0.05mm}
}

%% file: macros.tex
\renewcommand{\vec}[1]{\mathbf{#1}}

\newcommand{\bv}{\vec{b}}
\newcommand{\cv}{\vec{c}}

\newcommand{\sv}{\vec{s}}

\newcommand{\uv}{\vec{u}}

\newcommand{\xv}{\vec{x}}
\newcommand{\yv}{\vec{y}}

\newcommand{\Am}{\vec{A}}

\newcommand{\Dm}{\vec{D}}

\newcommand{\Lm}{\vec{L}}

\newcommand{\Pm}{\vec{P}}

\newcommand{\Um}{\vec{U}}

\newcommand{\FF}{\mathbb{F}}



%% file: tikz/blt.tikz
\begin{tikzpicture}[y=-.7cm,x=.7cm]

\tikzstyle{block} = [black, line width = 0.3mm];
\tikzstyle{solidline} = [line width = .3mm];
\tikzstyle{dottedline} = [dotted, line width = .3mm];

\draw[thick,block] (0,0) rectangle (7,7);
\draw[block] (0,0) rectangle (2,2);
\draw[block] (2,2) rectangle (3.5,3.5);
\draw[block] (5,5) rectangle (7,7);

\draw[loosely dotted, line width = .5mm] (4,4) -- (4.5,4.5);
\draw[dottedline] (2,3.5) -- (2,5);
\draw[dottedline] (3.5,3.5) -- (3.5,5);
\draw[dottedline] (3.5,5) -- (5,5);

\draw[solidline] (0,3.5) -- (2,3.5);
\draw[solidline] (0,5.0) -- (3.5,5.0);
\draw[solidline] (2,5.0) -- (2,7);
\draw[solidline] (3.5,5.0) -- (3.5,7);

\node[right] at (2,1){$ s_0 $} ; 
\node[below] at (1,2){$ s_0 $} ; 

\node[right] at (3.5,2.75){$ s_1 $} ; 
\node[below] at (2.75,3.5){$ s_1 $} ; 

\node[right] at (7,6.0){$ s_{m-1} $} ; 
\node[below] at (6.0,7){$ s_{m-1} $} ; 

\node[] at (5.5,1.5){\Huge$0$} ; 
\node[] at (1,1){$\Dm_{0,0}$} ; 
\node[] at (2.75,2.75){$\Dm_{1,1}$} ; 
\node[] at (1,2.75){$\Dm_{0,1}$} ; 
\node[] at (6.0,6.0){$\Dm_{m-1,m-1}$} ; 
\node[] at (1,6.0){$\Dm_{0,m-1}$} ; 
\node[] at (2.75,6.0){$\Dm_{1,m-1}$} ; 

\node[left, yshift=-.2cm] at (0,0){$ \gamma_0 = 0 $} ; 
\node[left, yshift=-.2cm] at (0,2){$ \gamma_1 = s_0 $} ; 
\node[left, yshift=-.2cm] at (0,5.0){$ \gamma_{m-1} $} ;

\end{tikzpicture}

%% file: tikz/multiplicity_chart_bhattacharyya.tikz
\begin{tikzpicture}
\begin{axis}[
width=\linewidth,
height=0.6\linewidth,
xmajorgrids,
yminorticks=true,
ymajorgrids,
ylabel={$|\operatorname{Aut}(\mathcal{C})|$},
xlabel={Design Erasure Probability $ \epsilon $},
ymode=log,
xmode=log,
mark size=1.5pt,
xmin=1e-2,
xmax=5e-1,
ymin=1e7,
extra y ticks ={268435456, 20972799094947840},
extra y tick labels={\color{lila}{\footnotesize\raisebox{4mm}{$ |\operatorname{LTA}(2,7)| $}}, \color{lila}{\footnotesize\raisebox{-4mm}{$ |\operatorname{GA}(2,7)| $}}},
extra tick style={major grid style=lila},
]

\addplot [color=mittelblau,solid, thick] 
table[row sep=newline, col sep=comma]{
1.000e-02,20972799094947840
2.557e-02,20972799094947840
2.659e-02,1775700541440
5.815e-02,1775700541440
6.047e-02,50734301184
1.487e-01,50734301184
1.546e-01,2415919104
2.199e-01,2415919104
2.287e-01,268435456
2.781e-01,268435456
2.891e-01,2415919104
5.000e-01,2415919104
};

\node[] at (2.5e-2,4e7) {\footnotesize $ \leftarrow $ ``RM-like''};
\node[] at (2.5e-1,4e7) {\footnotesize  ``Polar-like'' $ \rightarrow $};

\end{axis}

\end{tikzpicture}

%% file: tikz/multiplicity_chart_generators.tikz
\begin{tikzpicture}
\begin{axis}[
width=\linewidth,
height=0.6\linewidth,
xmajorgrids,
yminorticks=true,
ymajorgrids,
ylabel={$|\operatorname{Aut}(\mathcal{C})|$},
xlabel={Number of Generators $ |I_\mathrm{min}| $},
legend columns=1,
legend pos=north east,   
legend cell align={left},
legend image post style={mark indices={}},
ymode=log,
mark size=1.5pt,
xmin=0.8,
xmax=8.2,
ymin=1e7,
ymax=5e17,
extra y ticks ={268435456, 20972799094947840},
extra y tick labels={\color{lila}{\footnotesize\raisebox{4mm}{$ |\operatorname{LTA}(2,7)| $}}, \color{lila}{\footnotesize\raisebox{-4mm}{$ |\operatorname{GA}(2,7)| $}}},
extra tick style={major grid style=lila},
]

\addplot [color=rot,only marks,mark=x, thick, mark size=2.5] 
table[row sep=newline, col sep=comma]{%
1, 20972799094947840
2, 165140150353920
3, 7863816683520
4, 253671505920
5, 84557168640
6, 5637144576
7, 805306368
8, 268435456
};
\addlegendentry{\footnotesize Maximum};

\node[rot, anchor=west] (source) at (1.6, 1e17){\footnotesize{RM(3,7)}};
\node (destination) at (1, 20972799094947840){};
\draw[->, rot](source)--(destination);

\addplot [color=mittelblau,only marks,mark=o, thick, mark size=2.5] 
table[row sep=newline, col sep=comma]{%
1, 5286894690631680
2, 12943987514481
3, 99128835166
4, 5406423192
5, 1687767837
6, 676759811
7, 340018244
8, 268435456
};
\addlegendentry{\footnotesize Average};

\addplot [color=apfelgruen, only marks,mark=square, thick, mark size=2.5]
table[row sep=newline, col sep=comma]{%
1, 1775700541440
2, 805306368
3, 268435456
4, 268435456
5, 268435456
6, 268435456
7, 268435456
8, 268435456
};
\addlegendentry{\footnotesize Minimum};

\addplot [color=orange,only marks,mark=diamond, thick, mark size=2] 
table[row sep=newline, col sep=comma]{%
	6, 2415919104
};

\node[orange, anchor=west] (source) at (6.5, 1e10){\footnotesize{Bhat. @1dB}};
\node (destination) at (6, 2415919104){};
\draw[->, orange](source)--(destination);

\node[] at (2.5,4e7) {\footnotesize $ \leftarrow $ ``RM-like''};
\node[] at (6.5,4e7) {\footnotesize  ``Polar-like'' $ \rightarrow $};

\end{axis}

\end{tikzpicture}

%% file: tikz/polar_256_aut_sc.tikz
\begin{tikzpicture}
\begin{axis}[
width=\linewidth,
height=.7\linewidth,
grid style={dotted,anthrazit},
xmajorgrids,
yminorticks=true,
ymajorgrids,
legend columns=1,
legend pos=south west,   
legend cell align={left},
xlabel={$E_\mathrm{b}/N_0$ [dB]},
ylabel={BLER},
legend image post style={mark indices={}},
ymode=log,
mark size=1.5pt,
xmin=1,
xmax=4,
ymin=3e-5,
ymax=8e-01
]


\addplot[color=mittelblau,line width = 1pt, solid,mark=o, mark size=1.5pt, mark options={solid}]
table[col sep=comma]{
	1.00, 5.084e-01
	1.50, 2.797e-01
	2.00, 1.298e-01
	2.50, 4.566e-02
	3.00, 1.195e-02
	3.50, 2.677e-03
	4.00, 4.669e-04
};
\label{plot:bhatta256_sc}

\addplot[color=mittelblau,line width = 1pt, dashed,mark=o,mark size=1.5pt, mark options={solid}]
table[col sep=comma]{
	1.00, 5.233e-01
	1.50, 2.932e-01
	2.00, 1.317e-01
	2.50, 4.503e-02
	3.00, 1.193e-02
	3.50, 2.685e-03
	4.00, 4.936e-04
};
\label{plot:bhatta256_aut_sc8}

\addplot[color=mittelblau,line width = 1pt, dotted,mark=o,mark size=1.5pt, mark options={solid}]
table[col sep=comma]{
	1.00, 2.410e-01
	1.50, 9.172e-02
	2.00, 2.750e-02
	2.50, 6.935e-03
	3.00, 1.791e-03
	3.50, 5.030e-04
	4.00, 1.340e-04
};
\label{plot:bhatta256_scl8}


\addplot[color=apfelgruen,line width = 1pt, solid,mark=square, mark size=1.5pt, mark options={solid}]
table[col sep=comma]{
1.00, 7.473e-01
1.50, 5.693e-01
2.00, 3.587e-01
2.50, 1.829e-01
3.00, 7.198e-02
3.50, 2.359e-02
4.00, 5.621e-03
};
\label{plot:31_57_sc}

\addplot[color=apfelgruen,line width = 1pt, dashed,mark=square,mark size=1.5pt, mark options={solid}]
table[col sep=comma]{
1.00, 3.522e-01
1.50, 1.606e-01
2.00, 4.613e-02
2.50, 9.211e-03
3.00, 1.041e-03
3.50, 1.196e-04
};
\label{plot:31_57_aut_sc8}

\addplot[color=apfelgruen,line width = 1pt, dotted,mark=square,mark size=1.5pt, mark options={solid}]
table[col sep=comma]{
1.00, 4.018e-01
1.50, 1.892e-01
2.00, 6.315e-02
2.50, 1.439e-02
3.00, 2.128e-03
3.50, 1.630e-04
};
\label{plot:31_57_scl8}


\addplot[color=rot,line width = 1pt, solid,mark=x, mark size=2pt, mark options={solid}]
table[col sep=comma]{
1.00, 6.601e-01
1.50, 4.590e-01
2.00, 2.481e-01
2.50, 1.155e-01
3.00, 4.041e-02
3.50, 9.347e-03
4.00, 1.821e-03
};
\label{plot:31_99_sc}

\addplot[color=rot,line width = 1pt, dashed,mark=x,mark size=2pt, mark options={solid}]
table[col sep=comma]{
1.00, 3.926e-01
1.50, 2.163e-01
2.00, 1.097e-01
2.50, 4.441e-02
3.00, 1.638e-02
3.50, 4.829e-03
4.00, 8.942e-04
};
\label{plot:31_99_aut_sc8}

\addplot[color=rot,line width = 1pt, dotted,mark=x,mark size=2pt, mark options={solid}]
table[col sep=comma]{
1.00, 2.655e-01
1.50, 9.985e-02
2.00, 2.664e-02
2.50, 5.130e-03
3.00, 7.440e-04
3.50, 9.000e-05
};
\label{plot:31_99_scl8}

\coordinate (legend) at (axis description cs:0.02,.02);

\end{axis}

\matrix [
draw,
fill=white,
matrix of nodes,
align =left,
row sep = -3,
column sep = -5,
inner sep= 2,
anchor=south west,
font=\footnotesize,
mark options={solid}
] at (legend) {
	Design & SC & SCL-8 & Aut-8-SC\\
	Bhat. @1dB & \ref{plot:bhatta256_sc} & \ref{plot:bhatta256_scl8} & \ref{plot:bhatta256_aut_sc8}\\
	$ I_\mathrm{min}=\{31,99\} $ & \ref{plot:31_99_sc} & \ref{plot:31_99_scl8} & \ref{plot:31_99_aut_sc8}\\
	$ I_\mathrm{min}=\{31,57\} $ & \ref{plot:31_57_sc} & \ref{plot:31_57_scl8} & \ref{plot:31_57_aut_sc8}\\
};

\end{tikzpicture}

%% file: tikz/interlocked_cycles.tikz
\begin{tikzpicture}

\draw [dotted, fill=rot!20] plot [smooth cycle] %
coordinates {(1-.4,0-.4)(2+.4,0-.4)(2+.6,0+.6)(5-.6,0+.6)(5-.4,0-.4)
(5+.4,0-.4)(5,1)(1,1) };

\draw [dotted, fill=rot!20] plot [smooth cycle] %
coordinates {(6-.4,0-.4)(6+.4,0-.4)(6+.6,0+.6)(10-.6,0+.6)(10-.4,0-.4)
	(10+.4,0-.4)(10,1)(6,1) };

\draw [dotted, fill=rot!20] plot [smooth cycle] %
coordinates {(3-.4,0+.4)(3+.4,0+.4)(3+.6,0-.6)(7-.6,0-.6)(7-.4,0+.4)
	(7+.4,0+.4)(7,-1)(3,-1) };

\draw [dotted, fill=rot!20] plot [smooth cycle] %
coordinates {(8-.3,0-.4)(9+.3,0-.4)(9+.3,0+.4)(8-.3,0+.4) };

\foreach \x in {0,...,12} {
	\node[circle, draw, thick, minimum size=.75cm] at(\x,0) {\x};
};

\end{tikzpicture}

%% file: tikz/polar_128_aut_sc.tikz
\begin{tikzpicture}
\begin{axis}[
width=\linewidth,
height=.8\linewidth,
grid style={dotted,anthrazit},
xmajorgrids,
yminorticks=true,
ymajorgrids,
legend columns=1,
legend pos=south west,   
legend cell align={left},
xlabel={$E_\mathrm{b}/N_0$ [dB]},
ylabel={BLER},
legend image post style={mark indices={}},
ymode=log,
mark size=1.5pt,
xmin=1,
xmax=4,
ymin=1e-5,
ymax=8e-01
]


\addplot[color=mittelblau,line width = 1pt, solid,mark=o, mark size=1.5pt, mark options={solid}]
table[col sep=comma]{
1.00, 4.148e-01
1.50, 2.550e-01
2.00, 1.420e-01
2.50, 6.845e-02
3.00, 2.648e-02
3.50, 9.336e-03
4.00, 2.907e-03
};
\label{plot:bhatta128_sc}

\addplot[color=mittelblau,line width = 1pt, dashed,mark=o,mark size=1.5pt, mark options={solid}]
table[col sep=comma]{
1.00, 3.832e-01
1.50, 2.406e-01
2.00, 1.387e-01
2.50, 6.332e-02
3.00, 2.497e-02
3.50, 8.699e-03
4.00, 2.924e-03
};
\label{plot:bhatta128_aut_sc8}

\addplot[color=mittelblau,line width = 1pt, dotted,mark=o,mark size=1.5pt, mark options={solid}]
table[col sep=comma]{
1.00, 3.081e-01
1.50, 1.789e-01
2.00, 9.310e-02
2.50, 4.335e-02
3.00, 1.821e-02
3.50, 6.817e-03
4.00, 2.245e-03
};
\label{plot:bhatta128_scl8}

\addplot[color=mittelblau,line width = 1pt, dash dot,mark=o,mark size=1.5pt, mark options={solid}]
table[col sep=comma]{
1.00, 3.047e-01
1.50, 1.900e-01
2.00, 9.570e-02
2.50, 4.707e-02
3.00, 1.920e-02
3.50, 6.860e-03
4.00, 2.300e-03
};
\label{plot:bhatta128_osd}


\addplot[color=rot,line width = 1pt, solid,mark=square, mark size=1.5pt, mark options={solid}]
table[col sep=comma]{
1.00, 5.127e-01
1.50, 3.295e-01
2.00, 1.989e-01
2.50, 9.984e-02
3.00, 4.311e-02
3.50, 1.652e-02
4.00, 5.387e-03
4.50, 1.358e-03
};
\label{plot:27_56_sc}

\addplot[color=rot,line width = 1pt, dashed,mark=square,mark size=1.5pt, mark options={solid}]
table[col sep=comma]{
1.00, 3.102e-01
1.50, 1.819e-01
2.00, 7.001e-02
2.50, 2.530e-02
3.00, 1.000e-02
3.50, 3.035e-03
4.00, 9.351e-04
};
\label{plot:27_56_aut_sc8}

\addplot[color=rot,line width = 1pt, dotted,mark=square,mark size=1.5pt, mark options={solid}]
table[col sep=comma]{
1.00, 2.511e-01
1.50, 1.275e-01
2.00, 5.475e-02
2.50, 2.129e-02
3.00, 7.680e-03
3.50, 2.647e-03
4.00, 8.357e-04
};
\label{plot:27_56_scl8}

\addplot[color=rot,line width = 1pt, dash dot,mark=square,mark size=1.5pt, mark options={solid}]
table[col sep=comma]{
1.00, 2.242e-01
1.50, 1.111e-01
2.00, 4.923e-02
2.50, 2.007e-02
3.00, 7.278e-03
3.50, 2.635e-03
4.00, 8.645e-04
};
\label{plot:27_56_osd}


\addplot[color=apfelgruen,line width = 1pt, solid,mark=x, mark size=2pt, mark options={solid}]
table[col sep=comma]{
1.00, 5.714e-01
1.50, 4.320e-01
2.00, 2.846e-01
2.50, 1.541e-01
3.00, 7.188e-02
3.50, 2.812e-02
4.00, 8.225e-03
4.50, 1.780e-03
};
\label{plot:23_112_sc}

\addplot[color=apfelgruen,line width = 1pt, dashed,mark=x,mark size=2pt, mark options={solid}]
table[col sep=comma]{
1.00, 2.866e-01
1.50, 1.460e-01
2.00, 5.857e-02
2.50, 1.972e-02
3.00, 5.640e-03
3.50, 9.450e-04
4.00, 1.603e-04
};
\label{plot:23_112_aut_sc8}

\addplot[color=apfelgruen,line width = 1pt, dotted,mark=x,mark size=2pt, mark options={solid}]
table[col sep=comma]{
1.00, 2.420e-01
1.50, 1.150e-01
2.00, 4.233e-02
2.50, 1.202e-02
3.00, 2.647e-03
3.50, 4.830e-04
4.00, 9.00e-05
};
\label{plot:23_112_scl8}

\addplot[color=apfelgruen,line width = 1pt, dash dot,mark=x,mark size=2pt, mark options={solid}]
table[col sep=comma]{
1.00, 1.560e-01
1.50, 6.077e-02
2.00, 1.985e-02
2.50, 4.583e-03
3.00, 1.133e-03
3.50, 2.821e-04
4.00, 7.627e-05
};
\label{plot:23_112_osd}


\addplot[color=mittelgrau,line width = .7pt, solid,mark=diamond, mark size=2pt, mark options={solid}]
table[col sep=comma]{
1.00, 6.726e-01
1.50, 5.242e-01
2.00, 3.689e-01
2.50, 2.306e-01
3.00, 1.234e-01
3.50, 5.605e-02
4.00, 2.101e-02
4.50, 6.320e-03
5.00, 1.070e-03
5.50, 2.215e-04
6.00, 2.270e-05
};
\label{plot:rm_sc}

\addplot[color=mittelgrau,line width = .7pt, dashed,mark=diamond,mark size=2pt, mark options={solid}]
table[col sep=comma]{
1.00, 2.792e-01
1.50, 1.520e-01
2.00, 5.427e-02
2.50, 1.500e-02
3.00, 3.283e-03
3.50, 4.340e-04
4.00, 4.338e-05
4.50, 2.966e-06
};
\label{plot:rm_aut_sc8}

\addplot[color=mittelgrau,line width = .7pt, dotted,mark=diamond,mark size=2pt, mark options={solid}]
table[col sep=comma]{
1.00, 2.925e-01
1.50, 1.514e-01
2.00, 6.257e-02
2.50, 1.996e-02
3.00, 4.727e-03
3.50, 7.874e-04
4.00, 8.940e-05
};
\label{plot:rm_scl8}

\addplot[color=mittelgrau,line width = .7pt, dash dot,mark=diamond,mark size=2pt, mark options={solid}]
table[col sep=comma]{
0.00, 5.025e-01
0.50, 3.205e-01
1.00, 1.538e-01
1.50, 5.590e-02
2.00, 1.538e-02
2.50, 4.031e-03
3.00, 6.489e-04
3.50, 1.063e-04
4.00, 1.090e-05
};
\label{plot:rm_osd}

\coordinate (legend) at (axis description cs:0.02,.02);

\end{axis}

\matrix [
draw,
fill=white,
matrix of nodes,
align =left,
row sep = -3,
column sep = -5,
inner sep= 2,
anchor=south west,
font=\footnotesize,
mark options={solid}
] at (legend) {
	Design & SC  & SCL-8 & Aut-8-SC & OSD-4 \\
	Bhat. @1dB & \ref{plot:bhatta128_sc} & \ref{plot:bhatta128_scl8} & \ref{plot:bhatta128_aut_sc8} & \ref{plot:bhatta128_osd} \\
	$ I_\mathrm{min}=\{27,56\} $ & \ref{plot:27_56_sc}  & \ref{plot:27_56_scl8} & \ref{plot:27_56_aut_sc8} & \ref{plot:27_56_osd} \\
	$ I_\mathrm{min}=\{23,112\} $ & \ref{plot:23_112_sc} & \ref{plot:23_112_scl8} & \ref{plot:23_112_aut_sc8} & \ref{plot:23_112_osd} \\
	RM(3,7) & \ref{plot:rm_sc} & \ref{plot:rm_scl8} & \ref{plot:rm_aut_sc8} & \ref{plot:rm_osd} \\
};

\end{tikzpicture}